\documentclass[11pt,twoside,a4paper]{article}
\usepackage[]{url}
\usepackage{listings}
\usepackage{graphicx}
\usepackage{colortbl}
\usepackage{algorithm}
\usepackage{algorithmic}
\usepackage{theorem}

\usepackage{amssymb}
\usepackage{graphicx}
\usepackage{paralist}

\usepackage{url}

\begin{document}

\title{RESTful or RESTless -- \\
Current State of Today's Top Web APIs} 
\author{Frederik B\"ulthoff, Maria Maleshkova\\
AIFB, Karlsruhe Institute of Technology (KIT), Germany\\
frederik.buelthoff@student.kit.edu, maria.maleshkova@kit.edu}
\date{}
\maketitle

\begin{abstract}
Recent developments in the world of services on the Web show that both the number of available Web APIs as well as the applications built on top is constantly increasing. This trend is commonly attributed to the wide adoption of the REST architectural principles \cite{rest}. Still, the development of Web APIs is rather autonomous and it is up to the providers to decide how to implement, expose and describe the Web APIs. The individual implementations are then commonly documented in textual form as part of a webpage, showing a wide variety in terms of content, structure and level of detail. As a result, client application developers are forced to manually process and interpret the documentation. Before we can achieve a higher level of automation and can make any significant improvement to current practices and technologies, we need to reach a deeper understanding of their similarities and differences. Therefore, in this paper we present a thorough analysis of the most popular Web APIs through the examination of their documentation. We provide conclusions about common description forms, output types, usage of API parameters, invocation support, level of reusability, API granularity and authentication details. The collected data builds a solid foundation for identifying deficiencies and can be used as a basis for devising common standards and guidelines for Web API development.
\end{abstract}

\section{Introduction}
Recent developments in the world of services on the Web show that both the number of available Web APIs as well as the applications built on top is constantly increasing\footnote{\url{http://blog.programmableweb.com/2013/04/30/9000-apis-mobile-gets-serious/}}.
Often this proliferation of programmable interfaces that rely solely on the use of URIs, for both resource identification and interaction, and HTTP for message transmission, is attributed to the wide adoption of the REST architectural principles~\cite{rest}. In particular, Web APIs are characterised by their relative simplicity and their natural suitability for the Web, employing the same technology stack, and these characteristics are exploited by many Web sites like Facebook, Google, Flickr and Twitter who offer easy-to-use, public APIs that provide simple access to some of the resources they hold, thus enabling third-parties to combine and reuse heterogeneous data coming from diverse services in data-oriented service compositions called mashups~\cite{restmashups}.

Despite their popularity, currently there is no widely accepted understanding of what a Web API is. In fact, while the term 'Web Service' is quite clearly defined
\cite{III2012}, Web APIs still lack a broadly accepted definition. Currently the term \textit{Web API} has a general, sometimes even controversial, meaning and is used for depicting HTTP-based component interfaces, frequently being inconsistent about the specific technical and design underpinnings. This situation is undoubtedly driven by the fact that, as opposed to Web service technologies, work around Web APIs has evolved in a rather autonomous way and it is up to the providers to decide how they are going to expose the interface, how they are going to document them and what characteristics these documentations have. As a result, the majority of the Web APIs are described only in human-oriented documentation in textual form, as part of webpages, which is very diverse in terms of structure, content and level of details~\cite{restaq}. Therefore, currently developers have to manually search for suitable documentation, interpret the provided details and implement custom solutions, which are hardly reusable. Such an approach to using Web APIs is very time and effort consuming and will not scale in the context of the growing number of exposed interfaces. 

Before any significant impact and improvement can be made to current Web API practices and technologies, we need to reach a deeper understanding of these. This involves, for instance, figuring out how current APIs are developed and exposed, what kind of descriptions are available, how they are represented, how rich these descriptions are, etc. It is only then that we shall be able to clearly identify deficiencies and realise how we can overcome existing limitations, how much of the available know-how on Web services can be applied and in which manner. 
To this end, we present a thorough analysis over the most popular Web APIs in ProgrammableWeb directory\footnote{http://www.programmableweb.com}.

The remainder of this paper is structured as follows: 
Section 2, describes the methodology used for conducting our Web API study, while Section 3 gives the collected data and provides a discussion on identified correlations and trends. Section 4 presents an overview of existing work on analysing Web services and Section 5 presents future work and concludes the paper.

%\vspace{-1em}
\section{Survey Setup}
The survey was conducted by a single domain expert during December 2013 and January 2014. The dataset, which comprises 45 Web APIs\footnote{Amazon Product Advertising, Amazon S3, BitBucket, Azure (Blob Service), Balanced Payments, Bing Maps REST Services, Bitly, Box, del.icio.us, Disqus, DocuSign Enterprise, Dropbox (Core API), eBay (Shopping API), Etsy, Eventful, Facebook (Graph API), Flickr, Foursquare, Freebase (Search/Reconcile), Geonames, GitHub, Google Custom Search, Google Maps API Web Services, Google Places API, Groupon, Heroku, Instagram, Last.fm, LinkedIn, OpenStreetMap (Editing API), Panoramio, Paypal, Reddit, Salesforce, Tropo, Tumblr, Twilio, Twitpic, Twitter, Wikipedia/Mediawiki, Yahoo! BOSS, Yahoo! BOSS Geo, Yammer, Yelp, Youtube} in total, was primarily composed through the use of the ProgrammableWeb directory. This popular directory provides basic information about Web APIs in general as well as their use in mashups. The latter was used as a metric for the popularity of the Web API. Since we wanted to capture the Web API characteristics that developers are most frequently faced with, we mostly chose the analysed APIs by taking those with the most mashups. This metric is however biased towards older established Web APIs, for which reason we included a third of the entries at random and through other popularity measures\footnote{Alexa.com rank and number of tagged questions on StackOverflow.com}.

While ProgrammableWeb is considered the largest directory of its kind and, therefore, best suited for this task, the information itself turned out to be in some cases incomplete or out of date, which made some changes to the dataset necessary. These problems included Web APIs that had been discontinued or replaced by others. In addition, similar Web APIs from the same provider, such as the various maps related from Google and Bing, were grouped together as a single entry. As a result, we retained a dataset containing 45 Web APIs.

The survey was conducted by manually analyzing the documentation made available by the Web API providers. The features, which were taken under consideration can be grouped into six categories, which include general Web API information, URI use, HTTP use, input and output data, security and policies as well as common design decisions. The examined criteria were gained from the key architectural principles of REST, the use of the underlying HTTP protocol and from common challenges and design decisions of Web API providers. The results from the survey, as given in Section 3 can, therefore, be used as a basis for judging to what extent todays top Web APIs are actually RESTful. The presented categories contain the following features:

%\vspace{-0.5em}
\begin{enumerate}
\item \textbf{General Web API Information} -- the APIs size in terms of operations, availability of other protocols and interface descriptions and the type of functionality provided. 
\item \textbf{URL and Resource Links} -- the kind of design schema used in the URL of the Web API and the use of links between API resources.
\item \textbf{HTTP Use} -- the used HTTP methods and support for alternative HTTP methods, how update operations are implemented, if meaningful HTTP status codes are used in cases of failure and how caching is addressed.
\item \textbf{Input and Output Data} -- which mechanisms are used for the transmission of input data, what types of input are there and what kind of output formats can be expected.
\item \textbf{Security and Policies} -- are limitations on the degree of utilization posted and enforced, is authentication necessary and if yes, which authentication scheme is supported.
\item \textbf{Common Design Decision} -- how are versioning and the selection of the output format realized.
\end{enumerate} 
\vspace{-0.7em}

The procedure for gathering the data was straightforward. For each Web API the corresponding ProgrammableWeb webpage and the provider's documentation were opened and examined. 
The heterogeneous nature of both the media and the structure of the documentation, as well as various different ways of conveying the same information made any kind of automation of the process unfeasible. Furthermore, some cases of unclear or missing information made it necessary to perform some test interactions with the Web APIs. 

%\vspace{-1em}
\section{Survey Results}

In this section we describe the results that have been collected as part of the survey on Web APIs. The recorded features have been grouped into six categories, each of which addresses a different aspect of the Web APIs. 

%\vspace{-1em}
\paragraph{}
\textbf{General Web API Information.} 
Counting the number of operations supported by the Web API gives us some measure of its size and, therefore, complexity. This metric is easily attained for RPC-style APIs. In the case of resource-oriented or RESTful Web APIs each combination of a resource and a HTTP-verb was counted as an operation. 
The majority (62\%) of the entries in our dataset had between eleven and one hundred operations, with 38\% in the 11-50 and 24\% in the 51-100 range.
The remaining Web APIs were roughly equally divided into a group of smaller (less than 11 operations, 20\%) and larger (more than 100 operations, 18\%) ones. Only two entries (4\%) provided a single operation.

Only a small percentage (20\%) of the Web APIs provided the same service using alternative protocols. For example, Flickr is available through SOAP and through XML-RPC as well, next to their request format self-described as REST.
In most cases if an alternative was available, it had been declared as a legacy protocol, not guaranteed to be up to date in functionality and developers were urged to switch to its HTTP-based Web API equivalent. In most cases these were alternative protocols -- SOAP %\cite{soap} 
or XML-RPC implementations, and had existed before the introduction of their Web API counterparts.

Interface descriptions in a machine-readable format, which in contrast to the textual documentation targeted at humans, can be automatically processed were available for only five Web APIs (11\%) -- three using a custom format and one case of JSON Hyper-Schema respectively WSDL.
Links to related resources embedded in the response data of Web APIs, which is an alternative to interface descriptions were available in eight cases (18\%). This is necessary for fulfilling the HATEOAS constraint of REST architecture, which requires, that, instead of interacting through a fixed predefined interface the client of a RESTful Web API will transition through application states by following links embedded in the resource representations.

Easier integration of Web APIs into applications can be aided through extensive tooling support. This can either be done directly through Software Development Kits (SDK)s or by providing metadata on the use of the Web API through interface descriptions. We differentiate two types of SDKs -- those which are developed and maintained by the provider of the Web API themselves (available in 58\% of the cases) and those provided by third parties (available in 51\% of the cases), but still named and linked to from within official documentation. In total was at least one SDK available (either official or unofficial) for 76\% of the Web APIs. The most commonly supported platform or programming language was Ruby, closely followed by Python, PHP, Java\footnote{Java and Android SDKs were both counted as Java}, C\#\footnote{C\# and .NET SKDs were both counted as C\#}, JavaScript and Objective-C\footnote{Objective-C and iOS SDKs were both counted as Objective-C}. In total, we counted 19 different platforms or programming languages supported by official and 44 by unofficial SDKs.

We can draw two main conclusions based on the gathered data. First, once an HTTP-based Web API is made available, providers tend to abandon and move away from previous interaction protocol implementations, such as SOAP. Second, machine-interpretable interface description formats are rather an exception than a rule. Most providers still prefer to document APIs directly as part of webpages. 

%\vspace{-1.3em}
\paragraph{}
\textbf{URLs and Resource Links.}
It can be argued that for truly RESTful Web APIs that follow the principle of HATEOAS (Hypermedia as the Engine of Application State) the URL design is opaque because the user of the Web API will never have to construct URLs manually. Nonetheless, the design or structure of the URLs remains a good indicator for the type of the Web API. In addition, we will see that only a small percentage of the Web APIs under consideration aim to follow the HATEOAS principle. We differentiated between three main types of URL design, those that were structured around resources (resouce-oriented) -- those that focused on the operations (RPC-style) and those in between (mixed). The latter category contains cases in which some parts, for example search was built in an RPC-style while the rest was structured around resources. 
The data in Table \ref{TypeOfAPI} shows that the majority was resource-oriented, followed by those in RPC-style, with the smallest group being those sorted into the mixed category. 

The availability of resource links was previously presented as part of the analysis on interface descriptions. 
The data in Table \ref{ResourceLinks} incorporates that number in addition to two further use cases: Web APIs with self links include the URL of resources as part of their representation and pagination links provide the user of the Web API with precomposed URLs for paging through datasets. Both help reduce the complexity of using the Web APIs but were only available in 13\% of the analyzed Web APIs.

%\vspace{-1.5em}
\begin{table}
\parbox{.45\linewidth}{
	\caption{URL Design}
	\centering
	\begin{tabular}{lll}
	\hline\noalign{\smallskip}
	Description & Number & In \% \\
	\noalign{\smallskip}\hline\noalign{\smallskip}
	 RESTful & 21 &	47 \\
	 RPC & 15 & 33 \\
	 Hybrid & 9 & 20 \\
	\hline
	\end{tabular}
	\label{TypeOfAPI}
}
%\hfill
\parbox{.45\linewidth}{
	\caption{Resource Links}
	\centering
	\begin{tabular}{lll}
	\hline\noalign{\smallskip}
	Description & Number & In \% \\
	\noalign{\smallskip}\hline\noalign{\smallskip}
	 Used at all & 11 &	24 \\
	 \hline\noalign{\smallskip}  
	 Related Resources & 8 & 18\\
	 Self & 6 & 13\\
	 Pagination & 6 & 13\\
	\hline
	\end{tabular}
	\label{ResourceLinks}
}
\end{table}
%\vspace{-1em}

The data indicates that HATEOAS remains one of the most poorly supported constraints of the REST architecture with less than a fifth of the analyzed Web APIs providing links to related resources. A possible explanation is that HATEOAS signifies the largest departure from the previous approaches on Web Services, which heavily relied on predefined interfaces. Notable exceptions include PayPal and Github, which explicitly feature HATEOAS respectively hypermedia links prominently in their documentation. 

%\vspace{-1em}
\paragraph{}
\textbf{HTTP Use.}
As it is to be expected, the two  most commonly used HTTP verbs are GET and POST (see Table \ref{MethodSupport}), since both are used by resource-oriented and RPC-style Web APIs. The least popular verb is PATCH. Most Web APIs (58\%) that feature update functionality use PUT or PATCH while 30\% use POST. The remaining 12\% break the idempotency of the GET verb by misusing it for update operations.

In some cases the more uncommonly used HTTP verbs, such as PATCH, are not supported by existing tools and frameworks. Some Web API providers offer, therefore, functionality that allows users to swap out the originally requested HTTP verb with another one, usually POST. 
Table \ref{MethodOverride} shows that the most popular way for indicating the original verb is by using a query parameter in the URL of the request. Others simply make no difference between the verb used or allow the requested verb to be set in either a custom header or the URL path. In total, a method override was provided by 42\% of the Web APIs.

Error handling plays a large role in any application. How Web APIs present errors is therefore of particular importance. 71\% of the surveyed Web APIs reused the various predefined status codes of HTTP to indicate an error. In all of those cases the body of the HTTP response did contain further information.

One advantage of using Web APIs and subsequently HTTP is the built-in support for caching, for which only 27\% of the Web APIs explicitly stated their support. Further manual analysis via test invocations showed that an additional six Web APIs did indeed support caching without having documented it. 

%\vspace{-1.5em}
\begin{table}
\parbox{.45\linewidth}{
	\caption{Method Support}
	\centering
	\begin{tabular}{lll}
	\hline\noalign{\smallskip}
	Description & Number & In \% \\
	\noalign{\smallskip}\hline\noalign{\smallskip}
	GET	& 45 & 100 \\
	POST & 34 & 76 \\
	DELETE & 21 & 47 \\
	PUT & 17 & 38 \\
	HEAD & 6 & 13 \\
	PATCH & 3 &	7 \\
	\hline
	\end{tabular}
	\label{MethodSupport}
}
%\hfill
\parbox{.45\linewidth}{
	\caption{Method Override}
	\centering
	\begin{tabular}{lll}
	\hline\noalign{\smallskip}
	Description & Number & In \% \\
	\noalign{\smallskip}\hline\noalign{\smallskip}
	Override Supported & 14 & 42 \\
	\noalign{\smallskip}\hline\noalign{\smallskip}
	Query parameter & 6 & 43 \\
	Interchangeable & 3 & 21 \\
	Header & 3 & 21 \\
	URL path & 2 & 14 \\
	\hline
	\end{tabular}
	\label{MethodOverride}
}
\end{table}
%\vspace{-1em}

Web APIs, which build upon the REST architectural principles, should embrace the HTTP protocol\footnote{REST is not tied to HTTP, but HTTP it is the base for communication on the world wide web and thus the most popular protocol which REST is applied to.}. Adopting the various aspects of HTTP enables the reuse of know-how and best practices gained in making the Web the way it is today. 
One part of adopting HTTP, means using the status codes defined in the standard, especially those for indicating the various types of errors, which may occur. We found out that the majority of the Web APIs use standard error codes. In contrast, cache support is not widely present, even though it is a feature, which Web API providers can easily support using the built-in mechanisms of HTTP.

%\vspace{-0.5em}
\paragraph{}
\textbf{Input and Output Data. }
Using Web APIs means interacting with data. Most requests to Web API will incorporate some input, which can be transmitted in many ways. 
Table \ref{ParamSource} shows that the analyzed Web APIs use four different ways for sending the input, the most popular one being input transmitted as parameters in the query string of the request URL. Another popular transmittal technique encodes the input in the request body, often by using the standard form encoding used by HTML forms on web pages or one of the supported output formats, such as JSON or XML. Many APIs support more than one type input encoding, especially when the output format itself can also be freely chosen.

The input can further be differentiated into several types (see Table \ref{ParamType}). All Web APIs under considerations had at least one case in which an input parameter was optional and almost all featured required parameters. In most cases information on which parameters must be provided, which ones are optional and what their associated default values are, is only provided out-of-band in textual documentation. Building valid requests which feature the expected data therefore require careful consideration. Further complexity arises from the fact that most Web APIs incorporate input parameters that \begin{inparaenum}[i)]
	\item state a list or range of valid values
	\item expect data to be encoded using a specific standard (e.g. dates as ISO 8601).
	\item or are of complex nature (e.g. comma separated lists of values). \end{inparaenum}
%Further difficulties that were recorded while conducting the survey include deprecated parameters and those with dependencies (e.g. the user may supply either parameter A or B but not both).

%\vspace{-1.5em}
\begin{table}
\parbox{.45\linewidth}{
	\caption{Way of transmitting input}
	\centering
	\begin{tabular}{lll}
	\hline\noalign{\smallskip}
	Description & Number & In \% \\
	\noalign{\smallskip}\hline\noalign{\smallskip}
	Query & 43 & 96 \\
	Body & 34 & 76 \\
	Path & 25 & 56 \\
	Header & 8 & 18 \\
	\hline
	\end{tabular}
	\label{ParamSource}
}
%\hfill
\parbox{.45\linewidth}{
	\caption{Input datatypes}
	\centering
	\begin{tabular}{lll}
	\hline\noalign{\smallskip}
	Description & Number & In \% \\
	\noalign{\smallskip}\hline\noalign{\smallskip}
	Optional & 45 & 100\\
	Required & 44 & 98\\
	Alternative/Range & 43 & 96\\
	Specified & 40 & 89\\
	Complex & 38 & 84\\
	%Boolean & 41 & 91\\
	\hline
	\end{tabular}
	\label{ParamType}
}
\end{table}
%\vspace{-1em}

In contrast to SOAP and XML-RPC, which both use XML as the transport and output format, Web APIs most commonly (89\%) feature support for the more compact data representation format JSON\footnote{JavaScript Object Notation, an open standard for data interchange derived from the JavaScript language.}. Still, XML remains the second most used data format (58\%). 
The increasing popularity of JSON is further reflected by the fact that about half of the Web APIs using it, do not provide XML support. Less than a fifth also supported other formats\footnote{e.g. including CSV, RDF, YAML, PHP, RSS, Atom, WDDX or form encoded values}.
%, PHP, RSS, Atom, WDDX or form encoded values. 
All of the Web APIs supported either JSON or XML as their primary data representation format. Two Web APIs used their own custom data output format, which in both cases was based on JSON and provided a general structure for all responses.

%Truly formal and machine-readable type descriptions that describe the data structure and allow for further processing, such as automatic validation were provided by 20\% of the Web APIs, mostly by using standards such as JSON Schema and XML Schema followed by individual uses of DTD, Relax NG and an OWL-based ontology.

Our results show that preparing the input in the right format requires additional effort. Each request to a Web API demands careful consideration on which parameters to send, their format and ultimately how to transmit them. In addition, there is no general consensus in Web APIs on how to format even frequently occurring input such as date and time, thus requiring careful manual effort when doing service composition. On the other side of the request are JSON and XML the two main established data interchange formats for output, with JSON rapidly gaining on importance.

%There were no semantic descriptions of APIs at all.
%
%%%@ Conclusions:
% 6. Preparing the input in the right format requires additional effort
% 7. JSON and XML are the two main established formats for providing the output, with JSON gaining on importance. 
%\vspace{-0,5em}
\paragraph{}
\textbf{Security and Policies.}
%
%%Usage/Rate Limitations - text
%%Usage/Rate Headers - text
%%Authentication - table 
%%SSL Support -text
%
Security and policies or terms of use play an important part in the context of using Web APIs, since they determine the conditions and limits for actually accessing the APIs. Only two of the examined Web APIs did not use any kind of authentication. 
Roughly a third of the Web APIs require authentication only for operations, which perform data modification, but do not require authentication for reading resources. 
The most common way of identifying the client application or user is via an  API key (also called application id, client id or by similar terms) which is passed along with each request. 
Other, more secure approaches, are listed in Table \ref{Authentication}. The most common approach, used by two thirds of the Web APIs is OAuth %\cite{oauth1,oauth2} 
in its various protocol versions followed by the basic %\cite{basic} 
authentication protocol of HTTP. 
%%%@ Done: Which is the latter case, I do not understand, please explain 
In those Web APIs that used basic authentication, which sends the provided credentials in plaintext as part of a HTTP header,  this authentication method was almost always combined with SSL\footnote{Secure Sockets Layer, a cryptographic protocol which aims to provide communication security over the Internet}. In total, SSL was available for 91\% of the Web APIs and its use was mandatory for 41\% of those.

Most Web APIs (89\%) state and implement rate limitations, which restrict the number of invocations in a specific time frame. Consumers of the API have to follow these restrictions in order to prevent their requests or the entire application from being blocked. 
The limitations are either written down, as part of the documentation, or included with the general terms and conditions. A fifth of the Web APIs use custom HTTP headers to convey information about the remaining quota in every response, thus allowing the client application to dynamically adapt its use pattern.

%\vspace{-1.5em}
\begin{table}
\caption{Common Web API Authentication Approaches}
\centering
\begin{tabular}{lll}
\hline\noalign{\smallskip}
Authentication Mechanisms & Number & In \% \\
\noalign{\smallskip}\hline\noalign{\smallskip}	
OAuth 1.0 & 20 & 44 \\
OAuth 2.0 & 11 & 24 \\
Custom OAuth & 2 & 4 \\
HTTP Basic &	8 & 18 \\
Session &	5 & 11 \\
Custom HMAC & 3 & 7 \\
Other & 4 & 9 \\
\hline
\end{tabular}
\label{Authentication}
\end{table}
%\vspace{-1em}

We can conclude, that the majority of Web APIs use authentication in some form, requiring adopters of these services to both register their application in advance and tackle the individual authentication scheme used. Our results show that OAuth has the potential to emerge as universally adopted standard for authentication. Almost as common as authentication are limitations on the number of requests per time period that applications can send to Web APIs. 

%\vspace{-0,5em}
\paragraph{}
\textbf{Common Design Decisions.} 
The motivation behind versioning is that Web APIs may change over time and by explicitly distinguishing between versions, new releases will not break compatibility with older API clients. 
This issue was addressed by 73\% of the examined Web APIs. The most common technique, as shown in Table \ref{Versioning}, includes the API version as a prefix in the URL path. 
Further techniques include a custom HTTP header, standard content negotiation, the specification of versions in the body of the request and switching subdomains. The latter technique was used by Facebook to differentiate their deprecated REST API from their new Graph API.

The way of selecting the output format is another common design decision for Web APIs. Four different techniques (see Table \ref{FormatSelection})  were identified during the survey. 
In six cases did the Web APIs support more than one way of requesting a specific format. 
The two most common methods include specifying the format as part of the URL, either as part of the path or as a query parameter. The standard mechanism of HTTP for this purpose, content negotiation (also used for versioning purposes as seen above), was supported by six APIs, followed by the use of a custom HTTP header by two.

\begin{table}[H]
\caption{Common Web API Versioning Techniques}
\centering
\begin{tabular}{lll}
	\hline\noalign{\smallskip}
	Description & Number & In \% \\
	\noalign{\smallskip}\hline\noalign{\smallskip}
	Yes & 33 & 73\\
	No & 12 & 27\\
	\hline\noalign{\smallskip}	
	URL Path &	26 & 79\\
	Custom Header &	2 & 6\\
	Content-Negotiation	&	2 & 6\\
	Body & 2 & 6\\
	Subdomain change &	1 & 3\\
	\hline
\end{tabular}
\label{Versioning}
\end{table}

\begin{table}[H]
\caption{Representation Format Selection}
\centering
\begin{tabular}{lll}
	\hline\noalign{\smallskip}
	Description & Number & In \% \\
	\noalign{\smallskip}\hline\noalign{\smallskip}	
	Yes & 28 & 62\\	
	No & 12 &	27\\	
		\hline\noalign{\smallskip}	
	Path/File extension & 15 & 54\\	
	Query Parameter & 11 & 39\\	
	Content-Negotiation	& 6 & 21\\	
	Custom Header & 2 &	7\\	
	\hline
\end{tabular}
\label{FormatSelection}
\end{table}

Our results show that even though HTTP defines content negotiation using the accept header as the mechanism for representation format selection, is it only supported by a minority of Web APIs. Instead, most Web APIs allow the format to be specified in some way as part of the URL, which allows basic requests to be easily tested in a common web browser. For versioning, including the version identifier as part of the URL is by far the most popular technique. 

\vspace{-1em}
\section{Related Work}

The first study on the state of Web APIs was presented by \cite{maria} and features a comprehensive overview through the analysis of 222 Web APIs in 2010. 
While Maleshkova et al. aim to draw conclusions on the state of the entire world of  APIs on the Web, we focus on the most popular and common ones, substituting a larger dataset for more and other types of features. Another more recent study from 2012 was provided by Renzel et al. \cite{top20}, wherein the authors analyze a dataset of twenty Web APIs by a broad range of features, some of which were incorporated in our survey. Similar to our study, the dataset was gained by selecting top ranked entries from the ProgrammableWeb directory, using the number of mashups as the sorting criteria. The rather limited dataset and fast moving developments in the world of services on the web necessitate taking another look at the current state of Web APIs. Other older studies, devoted to investigating Web Services exist. The authors in \cite{li} provide a study on Web services but their data is restricted to only a few characteristics and a single source. 

\vspace{-1em}
\section{Conclusions and Future Work}

The results of our survey indicate that Web APIs feature a large amount of heterogeneity in their individual designs, ranging from cases, following the architectural style of REST and its constraints, such as HATEOAS, to those featuring a more RPC-like style. Common service tasks such as composition and invocation, therefore, require more manual effort to smooth over differences in implementations, compared to Web Services that follow a strict standard such as SOAP. Even though REST is only an architectural style, in contrast to a strict standard such as SOAP whose conformity can be validated, a stricter compliance with its guidelines and constraints would already significantly reduce friction in adopting Web APIs for more complex tasks such as the automation of composition and invocation. While some more readily understandable concepts such as using the HTTP verbs have gained widespread adoption, other concepts such as resource linking (HATEOAS) are hardly ever applied. For today's top Web APIs we, therefore, have to conclude that they most commonly remain RESTless.

This area of research has a lot of potential for further work. By building upon the data gained as part of this survey and the previous ones mentioned in the related work section, we could quantify the changes in Web API design over time and possibly gain insight over future developments. Another idea would be to take those parts of the REST principles that we have shown to be poorly applied and work on the problems surrounding their adoption.

%\textbf{Acknowledgments} The work presented in this paper is partially supported by EU funding under the project xLiMe (FP7 - 611346).

\vspace{-1em}
\bibliographystyle{splncs03}

\end{document}